\documentstyle[12pt,aaspp4]{article}

\newcommand{\Msun}{\mbox{ M}_{\odot}}
\newcommand{\gae}{\mathrel{>\kern-1.0em\lower0.9ex\hbox{$\sim$}}}
\newcommand{\kms}{km~s$^{-1}$}

\lefthead{Boroson et al.}
\righthead{Models of LMC X-4}

\begin{document}

\title{Testing Hydrodynamic Models of LMC X-4
with UV and X-ray Spectra}

\author{Bram Boroson and Timothy Kallman}
\affil{Goddard Space Flight Center, Greenbelt, MD 20771;
bboroson@falafel.gsfc.nasa.gov, tim@xstar.gsfc.nasa.gov}

\and

\author{John M. Blondin and Michael P. Owen}
\affil{Department of Physics, North
Carolina 
State University; blondin@tigger.physics.ncsu.edu,
owen@anubis.physics.ncsu.edu}

 


\begin{abstract}

We compare the predictions of hydrodynamic models of the LMC X-4 X-ray
binary system with observations of UV P~Cygni lines with the GHRS and STIS
spectrographs on the Hubble Space Telescope. The hydrodynamic model
determines density and velocity fields of the stellar wind,
wind-compressed disk, accretion stream, Keplerian accretion disk, and
accretion disk wind. We use a Monte Carlo code to determine the UV~P~Cygni
line profiles by simulating the radiative transfer of UV photons that
originate on the star and are scattered in the wind.  The qualitative
orbital variation predicted is similar to that observed, although the
model fails to reproduce the strong orbital asymmetry (the observed
absorption is strongest for $\phi>0.5$).  The model predicts a mid-eclipse
X-ray spectrum, due almost entirely to Compton scattering, with a factor~4
less flux than observed with ASCA. We discuss how the model may need to be
altered to explain the spectral variability of the system. \end{abstract}

\section{Introduction}

The mass transfer in High Mass X-ray binaries (HMXB) has proved to be
difficult to understand in spite of the simplicity of the first analytic
descriptions of accretion from a wind (Bondi \&\ Hoyle 1944). Problems
arise in the theory, for example, when X-ray feedback on the dynamics is
considered (MacGregor \&\ Vitello 1982; Ho \&\ Arons 1987).  The feedback
between the X-ray source and a stellar wind can be severe, as the mass
accretion rate of a stellar wind depends on $v_w^{-1} v_{\rm
rel}^{-3}=v_w^{-1} (v_w^2+v_x^2)^{-3/2}$, where $v_{\rm rel}$ is the
velocity of an
X-ray source with orbital velocity $v_x$ relative to a stellar wind with
velocity $v_w$. The X-rays can reduce $v_w$ by ionizing to higher stages
those ions responsible for the acceleration of the wind.

Even isolated stellar winds are complex. The line-scattering force
accelerating the wind depends on the wind velocity gradient.  This leads
to instabilities such as shocks (MacGregor, Hartmann, \&\ Raymond 1979;
Abbott 1980; Lucy \&\ White 1980; Owocki \&\ Rybicki 1984) and
nonlinearities which are difficult to model numerically (Owocki, Castor,
\&\ Rybicki 1988).

Near orbital phase $\phi=0.8$ many X-ray binaries show ``pre-eclipse
dips'' that may be caused when the gas stream that feeds the disk passes
over the line of sight.  Even systems such as 4U~1700-377 thought to
accrete primarily from a stellar wind rather than Roche lobe overflow
display increased X-ray absorption at phases $\phi\gae0.8$ (Haberl, White,
\&\ Kallman 1989, Heap \&\ Corcoran 1992).  This may be due to an
accretion wake, a bow shock that results when the wind encounters the
gravity of the compact object.

For both 4U~1700-37 and Vela X-1, increased absorption in optical lines is
seen at late orbital phases.  Accretion wakes are not expected to be as
large as the primary star, so are probably not responsible.  In Vela~X-1,
another structure caused by the interaction between the wind and the
compact object, a photionization wake, may cause the absorption (Kaper,
Hammerschlag-Hensberge, \&\ Zuiderwijk 1993). A photoionization wake has
also been invoked to explain the asymmetry of the X-ray eclipse in Vela
X-1 (Feldmeier et al. 1996).  This wind structure was first described by
Fransson \&\ Fabian (1980), and is seen in two-dimensional hydrodynamic
simulations (Blondin et al. 1991; Blondin 1994).  Photoionization wakes
result when the outflowing stellar wind in the shadow of the primary
enounters the slower-moving wind that has been stalled by X-ray
ionization.  Only near the eclipse ingress line does the moving X-ray
shadow of the star allow the fast-moving wind to continue to accelerate as
it plows into the stalled wind.

The observations of X-ray dips and eclipses, however, have not provided us
with information on the velocity field in the wind.  For this, the
UV~P~Cygni line profiles, formed by resonance scattering in the expanding
wind, have proved useful.  Hatchett \&\ McCray (1977) pointed out that the
P~Cygni lines in a HMXB would vary over the binary orbit in a
characteristic way. The ion responsible for the line would be removed from
a region surrounding the X-ray source, so that at $\phi=0.5$ the
blue-shifted absorption would be diminished, while at $\phi=0.0$ the
red-shifted emission would be diminished.  In general, the ``transparency
feature'' in the P~Cygni line would shift throughout the orbit depending
on the velocity field of the wind.  McCray et al. (1984) used this method
to infer the mass-loss rate and terminal wind velocity of Vela~X-1.  
Kaper et al. (1993) showed that the P~Cygni line variation in Vela~X-1
suggests that the wind velocity does not increase monotonically with
radius. A previous analysis of the P~Cygni lines of LMC~X-4 (Boroson et
al. 1999) attempted to infer the radial velocity law from the orbital line
variation.

While previous analyses of the wind velocity field from the
Hatchett-McCray effect have been semi-empirical, two- and
three-dimensional hydrodynamic simulations (Blondin et al. 1990, 1991;
Blondin 1994) have been increasing in sophistication.  Such models have
been constructed for particular HMXB systems, and have been compared in
some cases with X-ray observations (Blondin \&\ Woo 1995; Wojdowski,
Clark, \&\ Kallman 1999).  
The empirical analysis of the UV lines is hampered by the many degrees of
freedom required to describe the three-dimensional velocity and density
fields of the wind.  Greater physical insight can be provided if the
hydrodynamic simulations are used to predict P~Cygni profiles, which could
then be compared with the observations. This has not been previously
attempted, as most of the methods used to compute P~Cygni lines have
required the velocity and density fields to be spherically symmetric.  
Monte-Carlo simulation of P~Cygni profiles (Stevens 1993; Knigge, Woods,
\&\ Drew 1995), however, is not subject to this limitation. In this paper
we confront for the first time hydrodynamic models of stellar wind
disruption by a compact object with observations of the UV~P~Cygni lines.  
We also calculate the differential emission measure in the model of wind
and disk, and discuss the expected X-ray emission lines.

\section{The Hydrodynamic Model}

The hydrodynamical model used in this paper is a three-dimensional
time-dependent numerical simulation of an HMXB first presented by Owen \&
Blondin (1997).  The parameters of this model were chosen in accordance
with the observed parameters of LMC X-4 (see Table 1).  The high X-ray
luminosity of LMC X-4 is difficult to reconcile with wind accretion alone,
and the presence of a 30 day period in the X-ray light curve (Lang et al.
1981) and in the optical light curve (Heemskerk \& van Paradijs 1989)
suggests a precessing accretion disk fed by Roche lobe overflow.

\subsection{Numerical Method}

Our numerical model is an extension of previous  models (Blondin et
al.1990, Blondin et al.1991, Blondin  1994, Blondin \& Woo 1995) used
to study the dynamics of the circumstellar gas in HMXBs, and in
particular the disruption of the normal OB stellar wind by the compact
X-ray source.   These models are based on the time-dependent
hydrodynamics code VH-1, and include (1) a radiatively driven stellar
wind incorporated via the Sobolev approximation,  (2) the X-ray
photoionization of the wind as approximated by a cutoff of the
radiative driving force, and (3) the X-ray heating and radiative
cooling of the circumstellar gas.

The principal modification to this current model over that used in
Blondin \& Woo (1995) is the inclusion of the gravity of the compact
X-ray source.  This adds several features to the model, including tidal
distortion of the primary, Roche lobe overflow leading to the formation
of a tidal stream, the presence of an accretion disk, and a realistic
mass accretion rate.  In order to accurately model these features the
spherical numerical grid (188 by 63 by 201 zones in $r$, $\theta$, and
$\phi$) was structured to provide the highest spatial resolution near
the surface of the distorted primary and near the accreting compact
object.

The model is evolved only in one hemisphere, assuming reflection symmetry
about the equatorial plane.  In addition, a cone about the polar axis is
removed from the simulation to reduce the number of time steps needed in the
simulation.  The use of an explicit hydrodynamics code restricts the time step
to the smallest value of $\Delta x/v$ on the grid, where $\Delta x$ is the
width of a numerical zone, and $v$ is a characteristic velocity (either the
gas velocity or the sound speed).  On a spherical grid $\Delta x$ in the
$\phi$ direction is given by $r\sin\theta \,\Delta\phi$.  Therefore, as
$\theta\rightarrow 0$, the zone width $\Delta x\rightarrow 0$ and 
$\Delta t\rightarrow 0$.  Thus, to keep a reasonable value of $\Delta t$, we
limited the simulation domain to $\theta > 0.3\pi$.  

The numerical simulation was initialized with the surface of the primary star
just inside the critical potential surface, such that the effective potential
at the stellar surface was greater than the critical potential
by one part in $10^5$.  The boundary conditions on this
surface correspond to an isothermal atmosphere in hydrostatic equilibrium. 
At the stellar surface temperature of $4\times 10^4$ K, the scale height in
this atmosphere is $\sim 10^9$ cm.  From numerical experiments we found that
VH-1 could simulate a stable atmosphere with zone spacing as large as twice
the scale height; this is the value used in the present model.
The density at this boundary is set at $5\times 10^{-9}$ g cm$^{-3}$, a value
determined emperically to produce a stable stellar wind with the given 
parameters.  The parameters describing the radiative driving force 
($\alpha = 0.53$, $k=0.12$; see e.g., Castor, Abbott \& Klein 1975) 
were chosen based on the spectral type of the primary star. 
In the absence of a companion, this model 
would drive a spherical stellar wind with a mass loss rate of 
$\dot M_w \approx 2\times 10^{-7}\, {\rm M_\odot\, yr^{-1}}$ 
and a terminal wind velocity of 
$v_\infty \approx 2000$ km s$^{-1}$. 

The accretion-driven X-ray flux photo-ionizes, heats, and imparts
momentum to the circumstellar gas.   The X-ray luminosity was computed
self-consistently as a function of the mass accretion rate onto the
compact companion:  $L_x = 0.1 \dot Mc^2$.  This mass accretion rate,
$\dot M$, was computed by removing mass and energy from a block of
$3^3$ zones at the location of the companion (the zone size at this
location is $\Delta x = 2.5\times 10^9$ cm). However, the dense gas of
the tidal stream and accretion disk are expected to be highly optically
thick, and will shadow other regions of the binary system from this
intense X-ray flux. To account for this shadowing in the simplest
fashion, we computed the optical depth from the compact X-ray source to
each numerical zone, and reduced the X-ray flux by $e^{-\tau}$.  
The reduced flux was used to calculate the local photoionization parameter,
X-ray heating, and X-ray radiation pressure.

\subsection{Simulation Results}

The simulation was evolved for approximately one binary orbital
period.  This was sufficient time for the accretion disk to circularize
and for the circumstellar gas to reach a state of quasi-equilibrium.
By the end of the simulation there were five key features of the
circumstellar gas: a radiatively driven wind from the shadowed surface
of the primary star, a wind-compressed disk created by the rapid
rotation of the primary star, a tidal stream emanating from the inner
Lagrange point (L1), a Keplerian accretion disk around the compact
object, and an X-ray irradiation driven wind emanating from the inner
regions of the accretion disk.

The radiatively driven O-star wind fills a large fraction of the binary
system due to the strong shadowing effects of the accretion disk. This
normal stellar wind is marked by the presence of a wind-compressed disk
(Bjorkman \& Cassinelli 1993).  The density of the wind in the equatorial
plane is a factor of $\sim 20$ higher than at other latitudes.  This
effect is a direct result of assuming co-rotation between the primary star
and the binary system; a slower stellar rotation would diminish or remove
the wind-compressed disk.

The tidal stream is described remarkably well by the analysis of Lubow
\& Shu (1975).  The tidal stream leaves the L1 point with an angle of
$\sim 25^{\rm o}$, in good agreement with the analytic prediction.
The stream becomes supersonic only $10^{10}$ cm downstream, with
the centroid of the stream following a ballistic trajectory
until the stream impacts the disk.  The gas in the stream continues to 
accelerate, reaching a Mach number at impact of $\sim 16$.   

The width of the tidal stream ($\sim 7\times 10^{10}$ cm) is in good
agreement with Lubow \& Shu (1975); 70 \%  of the stream mass flux is
within their quoted width.  The stream cross section is consitently
oblate, with the stream width being about 4/3 of the stream height.  As
shown by Lubow \& Shu (1976), the vertical structure of the stream is
determined entirely by the foot points of the streamlines in the tidal
stream, and hence the stream has a scale height much larger than that
expected for hydrostatic equilibrium.  The density in the tidal stream,
and hence the mass transfer rate, is determined by the placement of the
stellar surface relative to the L1 point (an arbitrary boundary
condition in our model).  Both the width of, and the density in, the
tidal stream decrease as the stream accelerates toward the compact
object.

By the end of the simulation the accretion disk has relaxed into a
roughly steady-state configuration.  The accretion flow has fully
circularized and is nearly Keplerian, the vertical density profile
matches that of an isothermal steady-state disk, and the outer radius
is truncated at $\sim 8\times 10^{10}$ cm, corresponding to $\sim 70
\%$ of the Roche lobe of the compact object.  Note, however, that the
mass flux through the disk is not in equilibrium - even if such an
equilibrium were possible in this numerical model, it would take far
too long to reach.

The outer edge of the disk has a scale height of $\sim 2\times 10^{9}$
cm, although the disk is somewhat thinner on the side nearest the
primary star and fatter on the far side.  This scale height is
consistent with that of an isothermal accretion disk in steady state
(the sound speed of the cold disk gas in our model is 
$c_s = 2.136\times 10^6$ cm s$^{-1}$).  The width of the tidal stream in
the vicinity of the disk edge is noticeably wider, with a scale height
of order $\sim 3\times 10^{9}$ cm.  As a result, a significant fraction of
the tidal stream gas is not stopped at the disk edge, but rather flows 
over and under the disk along its surface.  This stream overflow ultimately
merges with the accretion disk in the vicinity of the point of closest
approach: the location at which a ballistic trajectory from the L1 point
would reach the closest to the compact companion.

In addition to the impact by the tidal stream, the outer edge of the
disk is continually ablated by the strong stellar wind from the primary
star.  Note that this equatorial wind is enhanced by the wind
compressed disk effect, and it is not directly affected by the X-ray
source since it is shadowed by the accretion disk.  As a result, the
equatorial wind is strong enough to ablate a significant mass flux off
the edge of the disk, particularly on the back side as seen in Figure~3.
Furthermore, the ram pressure of this equatorial wind inflates the
outer edge of the disk, creating a small outer rim that extends higher
than the rest of the disk.

Although the outer regions of the disk are sufficiently resolved by our
numerical grid, the inner regions of the disk are not.  The scale height
of a steady isothermal disk at a disk radius of $1.5\times 10^{10}$ cm
should be $\sim 2.6\times 10^8$ cm, but the limited resolution of our
numerical model limits the disk thickness to values approximately
three times larger.  This limitation has direct bearing on
the results of this paper in that this poor numerical resolution effectively
limits (through the high opacity of the disk material) the cone of X-rays
to a smaller opening angle than if the disk were adequately resolved at
small disk radii.

The intense X-ray luminosity of the accreting compact companion can
influence the circumstellar gas in many ways, but in this numerical model
the dominant effect is a strong bipolar wind driven off the inner region
of the accretion disk via X-ray irradiation.  The dense, cool gas on the
surface of the accretion disk is heated by the incident X-ray flux to
several million Kelvins.  The ensuing high thermal pressure drives a
strong wind with velocities reaching 2000 km s$^{-1}$.  This irradiation
is so intense in this numerical model that the X-ray flux striking the
inner disk edge completely ablates the inner disk, leaving a substantial
hole. As seen in Figure~4, this disk wind is roughly
spherical.  This has the interesting effect of producing a large volume of
X-ray heated gas with a roughly constant ionization parameter. Since both
the gas density and X-ray flux are decreasing as the square of the
distance from the X-ray source, the ionization parameter in this gas
remains at a constant value.

Because the poor numerical resolution in the inner accretion disk
affects the vertical structure of the disk, the conservation of angular
momentum in the disk gas, and the thermal transition of the X-ray
heated gas, the geometry and mass flux in this bipolar wind is very
uncertain.  Nonetheless, it is clear from both simple order of
magnitude estimates and from detailed calculations (de Kool \&
Wickramasinghe 1999) that the intense X-ray flux present in X-ray
binaries will drive a substantial mass loss from the accretion disk.

Due to the shadowing effects of the accretion disk, 
the effects of X-ray photoionization are restricted to the bipolar disk wind
in this model.
Gas outside of the disk wind is shielded from the X-ray source, and as a result,
there is no independent photoionization wake visible in this model.  Instead,
there is a relatively dense wake of circumstellar gas piled up along the leading
surface separating the disk wind and the stellar wind.  This modest wake, with
a density contrast of $\sim 4$, can be seen in the horizontal slices 
shown in Figure~5.

As seen in Blondin \& Woo (1995), the radiatively-driven wind of the
primary star is replaced by a thermally driven wind on the X-ray
irradiated face of the primary.  The high X-ray flux on the irradiated
surface is strong enough to supress the radiative driving force and quench
the normal radiatively driven stellar wind.  Instead, the X-ray flux heats
the outer atmosphere of the star up to X-ray temperatures, and the ensuing
high thermal pressure drives an outflow off the stellar surface.  In this
particular model, however, the thermally-driven stellar wind is
effectively snuffed by the disk wind.  The momentum in the disk wind is
sufficiently high that it stops the stellar wind before it can reach a
sonic point.  This subsonic outflow is then redirected up over the top of
the star along with the supersonic disk wind.  This dynamical interaction
between star and disk winds can be seen in Figure~3.

\section{Monte Carlo Simulation of P~Cygni Profiles}

Because the density and velocity fields $\rho(r,\theta,\phi)$ and 
${\bf v}(r,\theta,\phi)$ given
by the hydrodynamic model
are not spherically symmetric, we cannot use an escape probability method
(Castor 1970; Castor \&\ Lamers 1979) or the Sobolev with Exact
Integration method (Lamers, Cerruti-Sola, \&\ Perinotto 1987) to compute
the P~Cygni line profiles.  The lines we consider are doublets separated
by wavelengths less than the terminal wind velocity.  Thus photons emitted
by one doublet component can be scattered by the other, coupling different
regions of the wind. To circumvent the difficulty of solving the radiative
transfer integrals in this situation, we apply a Monte-Carlo method
similar to that of Knigge, Woods, \&\ Drew (1995).  Unique features of our
method are discussed in Appendix~A.  We show tests of our program in the
Appendix~B.

\subsection{Method}

We start the simulation by following a photon emitted by a random point on
the star's surface.
(The emission from the accretion disk should be dominated by emission from
the primary.  For the UV continuum, the disk contributes only
$\approx8$\%).
The random point is chosen with a weighting that takes into account
the gravity darkening
appropriate for a high-mass star with a radiative, rather than convective,
atmosphere.  The photon's initial direction is chosen randomly, but for
the frequency $\nu$, we step through a range centered on the rest
frequency $\nu_0$ with a width given by the maximum wind velocity.  The
details of the simulation are generally similar to those used by Knigge,
et al. (1995).  Because the wind velocity is not smooth or monotonic down
to scales
on the order of the grid spacing of the hydrodynamic simulation, we cannot
solve exactly for a resonant
point along the photon's path in which $\nu=\nu_0 (1+{\bf v_{\rm
wind}}\cdot {\bf n_{\rm photon}}/c)$.  To work around this, we allow a
microturbulent velocity within the wind. (For a treatment of
microturbulence using the Sobolev approximation for the source function in
a spherically symmetric wind, see 
Lamers, Cerruti-Sola, \&\ Perinotto
1987).  In addition to the practical benefit of allowing substantial
optical depths at nonresonant points, allowing microturbulence may also
simulate turbulent motions on a scale smaller than that of our numeric
grid.  The value of the
microturbulent velocity that we chose based on the thermal velocity and
the deviation in velocity among neighboring grid points is 4\%\ of the
terminal velocity.  Such a turbulent velocity
does not cause a dramatic change in the P~Cygni profiles of
symmetric winds.

\subsection{Auxiliary Assumptions}

In order to predict P~Cygni profiles from the density and velocity fields
computed by the hydrodynamic simulation, we need to make several auxiliary
assumptions.

First, we must assume abundances for Carbon and Nitrogen; these are
given in
Table~1.  Although metal abundances in the LMC are generally 20\%\ of
galactic abundances, we have chosen enhanced abundances because the
observed \ion{N}{5} and \ion{C}{4} lines are much stronger than expected
from a star of the spectral type of LMC~X-4.  As we will see in
\S\ref{sec:results}, this assumption has negligible effect on our
conclusions.

Second, we must determine the ionization stage of the wind both in the
presence and absence of X-ray illumination from the compact object.  In
the absence of X-ray illumination, we use an empirical relation
\begin{equation} \label{eqn:empirical} \tau_{\rm E}=T(1+\gamma)r^{-\gamma}
\end{equation} for the radial optical depth $\tau_{\rm E}$.  This
parameterization is often used to fit OB star spectra (Castor \&\
Lamers 1979). We used values of $T$ and $\gamma$ that we used in Boroson
et al. (1999) to fit the lines observed with the GHRS (see Table~1). In
view of the deep absorption troughs seen at $\phi=0.08$ and $\phi=0.9$
(atypical for an O7-9III-V star), we allowed an additional source of
ionization.  LMC~X-4 is known to show a residual X-ray flux during eclipse
that is $\approx1$\%\ of its maximum level (Woo et al. 1995), presumably
due to Thompson scattering of X-rays in the wind.  We determined the
ionization stage of the wind in response to the scattered X-rays using the
XSTAR photoionization code (Kallman et al. 1996).  Using the observed
X-ray spectrum of LMC~X-4 diluted to 1\%\ of the total luminosity, we
found the fraction of \ion{N}{5} and \ion{C}{4} for various values of the
ionization parameter $\xi\equiv L_{\rm x} / n_{\rm e} r^2$, where $r$ is
the distance from the X-ray source to the point in the wind, $L_{\rm x}$
is the X-ray luminosity, and $n_{\rm e}$ is the electron density.  The
optical depth then follows from the Sobolev approximation:
\begin{equation} \label{eqn:sobolev} \tau_{\rm S}=(\pi e^2/mc)f \lambda_0
n_{i,j} (dv/dr)^{-1} \end{equation} where $e$ is the electron charge, $m$
is the electron mass, $\lambda_0$ is the rest wavelength of the
transition, $f$ is the oscillator strength, $n_{i,j}=(n_{\rm e}/\mu) a_i
g_j$ is the number density of element $i$ in ionization stage $j$,
$\mu=1.2$ is the ratio of the mean atomic mass to that of hydrogen, and
$dv/dr$ is the wind velocity derivative in the direction of the photon's
path.  For the optical depth in the wind we then choose whichever of
$\tau_{\rm E}$ or $\tau_{\rm S}$ is greater.

In the presence of X-ray illumination (outside the
X-ray shadow of the normal star), we again compute
$\tau_{\rm E}$ and $\tau_{\rm S}$, but now we compute $\tau_{\rm S}$ using
the ion fractions given by X-ray illumination by the full luminosity of
LMC~X-4 (self-consistently calculated from the accretion rate of the
simulation to be $2\times10^{38}$~erg~s$^{-1}$). We now use
whichever of $\tau_{\rm E}$ or $\tau_{\rm S}$ is less.

As the simulation includes dense areas of the wind near the stellar
photosphere, the gas stream, and the accretion disk, collisional
de-excitation can prevent photons from scattering.  This causes
the absorption lines that we modelled empirically with
Gaussians in Boroson et al. (1999).  Our treatment of the stellar
absorption (see Appendix~A) is neccessarily approximate, as the
hydrodynamic simulation
does not fully resolve the stellar atmosphere.  The predictions of the
line profiles at low velocity, where the stellar absorption is strong,
should therefore be considered approximate.

\subsection{Limitations of the Model}

The hydrodynamic model was constructed in order to study the Roche lobe
overflow and formation of the accretion disk.  For these purposes, the
simulation did not need to extend beyond $2.5$
stellar radii.  We have not attempted to extrapolate the wind density and
velocity fields beyond $2.5 R_*$.  For a standard $\beta=0.8$ stellar wind
velocity law, that is 
\begin{equation}
\label{eqn:beta}
v(r)=v_\infty(1-R_*/r)^\beta,
\end{equation}
the wind only
reaches 66\%\ of its terminal velocity at the edge of the grid.  

The region about the poles of the normal star was excluded
from the simulation; the grid only extended from spherical coordinate
$\phi=52^\circ$ to $\phi=90^\circ$.  We simply interpolated the density
and temperature fields across the polar region; this should 
accurate enough for our current purposes,
as this region does not include the accretion wake, photoionization wake,
stream, or disk.  
The simulation only includes the region
above the orbital plane.  When a photon passes below this plane in our
simulation, it encounters density and velocity fields that are 
reflections of those above the plane.  
 
\section{Results\label{sec:results}}

\subsection{Orbital Variation of P Cygni Lines}

We compare the results of the simulation with the
\ion{N}{5}$\lambda\lambda1238,1242$ and
\ion{C}{4}$\lambda\lambda1548,1550$ lines observed with the GHRS and STIS
aboard the Hubble Space Telescope (Figures~\ref{fig:n5frame1} and
\ref{fig:c4frame1}).  The model spectra have errors in each 60~\kms bin
(determined from photon counting Poisson statistics of the Monte~Carlo
simulation) of $\approx5$\%\ of the flux.  For the STIS spectra, the
errors are $\approx1$\%\ in each 30~\kms bin.

The simulations reproduce qualitatively the observed Hatchett-McCray
effect. The absorption of the \ion{N}{5} and
\ion{C}{4} lines is reduced at $\phi=0.5$.  
In agreement with observations, the simulations predict some \ion{N}{5}
absorption remaining at
$\phi=0.5$.  This results from a combination of true absorption
(collisional de-excitation) and scattering from the photosphere and dense
regions
of the wind that are not fully ionized even when exposed to X-rays.
  
The orbital variability depends on the ratio $L_{38}/\dot{M_{-6}}$, where
$L_{38}$ is the X-ray luminosity in units of 10$^{38}$~erg~s$^{-1}$ and
$\dot{M_{-6}}$ is the mass-loss rate in units of $10^{-6}\Msun$~yr$^{-1}$.
In Boroson et al. (1999), we fit the P~Cygni lines
observed with the GHRS to find $L_{38}/\dot{M_{-6}}=0.26\pm0.01$ for
a homogeneous wind or $L_{38}/\dot{M_{-6}}=0.7\pm0.4$ for
a two-component wind with a density contrast $700\pm200$. 
In the hydrodynamic simulation, $L_{38}$ is fixed at 2, and 
$\dot{M_{-6}}=0.1-0.2$ (similar to the value found by Woo et al. 1995).
However, we note that the simulated wind
is highly inhomogeneous.  While 50\%\ of the wind volume has a local
$\dot{M_{-6}}=0.05-0.2$ (determined from continuity), 50\%\ of the
wind mass is in clumps denser by factors $\approx2000$.  These clumps are
found mostly in the wind-compressed disk.  

We investigated whether observable features in the P~Cygni profiles could
be traced to structures predicted by the hydrodynamic simulation. To do
this, we simulated the P~Cygni lines from a stellar wind first with a
spherically symmetric $v$ field given by Equation~\ref{eqn:beta} and the
same density field given by the simulation, and then with both a
spherically symmetric velocity field and a spherically symmetric density
field given by the continuity equation with
$\dot{M}=1.5\times10^{-7}\Msun$~yr$^{-1}$.  

The model predicts slightly greater absorption at $\phi=0.3$ than at
$\phi=0.7$ (Figure~\ref{fig:asym}) as a result of the aspherical wind
density.  The enhancement of the red-shifted emission at $\phi=0.7$ over
$\phi=0.3$ (for both \ion{N}{5} and \ion{C}{4}) results from the detailed
velocity field of the hydrodynamic simulation (not merely the effect of
the binary rotation).  For a wind with spherical symmetry and homogeneous
density, the ratio of the apparent optical depths in the absorption at the
blue doublet component to the absorption at the red doublet component is
$2:1$. In the inhomogeneous wind of the simulation, partial covering of
the star by dense clumps allows the doublet ratio to approach a
$\approx3:2$ ratio (determined from the profiles given by our Monte Carlo
calculation) even when the apparent optical depth $\tau\gg1$.  By the
apparent optical depth we mean $-\log F_\lambda$, where $F_\lambda$ is the
continuum-normalized flux at wavelength $\lambda$.  This is only an
apparent optical depth because the flux at $\lambda$ is the sum of fluxes
from different sightlines with different optical depths.

There are no signatures of the disk wind seen in
Figure~4 in the P~Cygni
profiles, as this wind is ionized to stages higher
than \ion{N}{5} and \ion{C}{4} by the unabsorbed
X-rays escaping perpindicular to the disk plane.

Some discrepancies between the data and model result from
the radial cutoff ($2.5 R_*$) and the terminal velocity
($2000$~km~s$^{-1}$) used in the simulation
($v_\infty=1150$~km~s$^{-1}$ fits the data better).
For example, the maximum observed velocity is $v_{\rm
max}=v_\infty(1-1/r_{\rm max})^\beta$.  Although the simulation
predicts the observed $v_{\rm max}$ at $\phi=0.08$, this results from too
large a value for $v_\infty$ and too small a value for $r_{\rm max}$
(which should approach infinity).
At $\phi=0.2-0.3$, when the X-rays ionize
the wind at large radii, the simulation predicts absorption in the wind
at too high a velocity.  For the \ion{C}{4} line, using too large
a value for $v_\infty$ results in a merging of the two doublet components.

The simulated profiles do {\it not} show the orbital asymmetry seen in IUE
equivalent widths (Boroson et al. 1999) or in the HST STIS spectra (more
absorption is seen for $0.5<\phi<1.0$ than for $0.0<\phi<0.5$).  This is
shown clearly for the case of \ion{N}{5} in Figure~\ref{fig:asym}.
Although we have made two assumptions designed to increase the ion
fraction of \ion{N}{5} (that the N abundance is not depleted as it is for
many LMC stars, and that scattered X-rays can increase the ion fraction),
we still find that the model does not reproduce the saturated \ion{N}{5}
absorption lines at $\phi=0.90$.  In order to produce saturated
absorption, the absorbing material must cover the entire stellar surface.

The presence of a \ion{He}{2} ionization front (Masai 1984) may invalidate
our use of simple scaling with $\xi$.  That is, at distances far enough
from the X-ray source, a large fraction of He (which has high
abundance) is in the form \ion{He}{2}, and is efficient at absorbing 
X-rays with $E>55$~eV.  The potential for ionizing \ion{N}{5} to higher
stages is 98~eV.  X-ray ionization would not be efficient at
ionizing \ion{N}{5} to higher stages if 98~eV photons are absorbed by
\ion{He}{2}.

A full treatment of the optically-thick ionization balance is beyond the
scope of this paper.  We point out, however, that when we apply
Equation~15 of Masai (1984) to the wind in LMC~X-4, we find that the
\ion{He}{2} ionization front would be at about 10~stellar radii from the
neutron star, outside of the grid of the hydrodynamical simulation.  In
the LMC~X-4 system, the X-ray luminosity is higher and the orbit is more
compact than for Vela~X-1, for which Masai's analysis may be more
appropriate.  This assumes a smooth wind with parameters given by the
spatially averaged values of the simulation.  Absorption by \ion{He}{2} in
an inhomogeneous wind may in general be more important.  
Most of the inhomogeneity in the present simulation is in the
wind-compressed disk
about the normal star.  We have found that the X-ray ionization effect
is mostly confined to the ionization of the accretion disk wind, as
elsewhere the wind is in the shadow of either the disk or the star.  The
disk wind should have a nearly constant ionization parameter $\log \xi>5$
so that He\,II should not cause much absorption within this region.
Although an ionization front
may decrease the effects of X-ray ionization on \ion{N}{5} and \ion{C}{4},
it is unlikely to do so in a manner asymmetric enough to cause the phase
asymmetry in the P~Cygni profiles (Figure~\ref{fig:asym}).

\subsection{Random Variability of P Cygni Lines}

We also simulated the P~Cygni line variability using density and velocity
fields given by a second frame of the hydrodynamic simulation, to get a
sense for how much P~Cygni variability is to be
expected from the ``weather'' of the system.  The time interval
between the two frames, 3300~seconds, is approximately the same
as the length of an HST orbit, and approximately the flow time across
one stellar radius.

The result of the comparison of P~Cygni profiles, shown in in
Figure~\ref{fig:difs}, shows that the random
variability does not exceed that expected from our simulated photon
statistics, $\approx5$\%\ of the flux in each 60~km~s$^{-1}$ wavelength
bin.  Thus if these simulations are representative, it should be safe to
assume that
variability in the line profiles throughout a binary orbit is a reflection
of changes in the orientation of the wind and not random variability
of the density and velocity fields.

\subsection{X-ray Spectra of the Wind}

From the density, temperature, and ionization in the hydrodynamic
model, we can also predict the contribution of the wind to the X-ray
spectrum.  This should include recombination and fluorescent lines
from the X-ray illuminated wind,
Compton scattering of the X-ray continuum from the neutron star,
and thermal bremsstrahlung emitted by the hot wind.  Following the
method of Wojdowski et al. (1999), we determine the recombination
and fluorescent spectra at each point in the wind from the local
$\xi$, the sum of a blackbody and a cut-off power law approximation to
the LMC X-4
X-ray spectrum, and
the XSTAR code. For the bremsstrahlung spectrum, we
use the temperature field of the hydrodynamic model instead of the
temperature calculated from XSTAR, as the wind can also be heated
by shocks.  We propagate the wind spectrum along the line of sight,
and allow absorption by the ionized wind and by an interstellar column
density of neutral gas $N_H=10^{21}$~cm$^{-2}$.  We assume a distance
to LMC~X-4 of 50~kpc, and a system inclination of 65$^\circ$.

We show the simulated mid-eclipse spectrum, its separate components, and
the observed ASCA spectrum from the campaign reported in Vrtilek et al.
(1997) in Figure~\ref{fig:windemit}.  The predicted spectrum, which we
have folded with the ASCA response, is due
almost entirely to Compton scattering.  If we fit the observed spectrum
with the model with arbitrary normalization, we find the observed
spectrum is a factor of~4 brighter than that predicted.  
We also
computed the wind spectra at $\phi=0.25,0.50,0.75$, but found no strong
line features.

Recombination and fluorescent lines provide a negligible contribution
because the disk shadows much of the wind.  Wojdowski et al. (1999) found
that the hydrodynamic simulation of the wind in SMC~X-1 predicts {\it
greater} recombination flux than is observed, and suggested this could
result from a low metal abundance ($Z<3\times10^{-2}$).  Wojdowski et al.
(1999) did not take into account the absorption of X-rays from the central
source to the wind.  No accretion disk was present in the SMC~X-1
simulation, although the high X-ray luminosity and the long-term period
are evidence that one exists.  The under-prediction of X-rays in
mid-eclipse by the LMC~X-4 simulation could result from too thick an
accretion disk.



\section{Future Work}

Some of the detailed predictions of the hydrodynamic model do not match
observations. In particular, the observed absorption troughs are deeper
than predicted by our simulations at $\phi=0.68,0.90$, and are deeper than
observed at $\phi^\prime=1-\phi$.

We plan to revise the boundary conditions of the simulation to see if the
observed line features can be reproduced.  The first adjustments to be
made should be minor adjustments to the current model: we know, for
example, that the terminal velocity of the wind should be closer to
1000~\kms than 2000~\kms.  The mass-loss rate may also be higher than the
$1-2\times10^{-7}\Msun$~yr$^{-1}$ used in the simulation.  Both of the
parameters depend on the details of the line-scattering responsible for
the wind acceleration, parameterized by $\kappa$ and $\alpha$ in Castor,
Abbott, \&\ Klein (1975).  
The values of $\kappa$ and $\alpha$ for the
O~star may differ from that assumed for the simulation because of
the different abundances in the LMC and in the Galaxy, or because of the
evolutionary status of the star.

If the mass-loss rate $\dot{M}$ were higher than that assumed in the
simulation, the P~Cygni absorption would be greater.  Two effects combine
to increase $\tau$ in Equation~\ref{eqn:sobolev}.  With a higher value of
$\dot{M}$, $n_{i,j}=(n_{\rm e}/\mu) a_i g_j$ is increased both because
$n_{\rm e}$ is increased on average (by conservation of mass) but also
because $g_j$, the fraction of ions in the absorbing stage is increased as
well.  However, the effects of an altered $\dot{M}$ on the hydrodynamics
remain to be investigated.

One possible explanation for the enhanced absorption at $\phi>0.5$ is that
a photoionization wake is present.  The simulations show a similar wake at
high stellar latitudes (Figure~5), but the wake does not
wind around the star to
cause absorption at $\phi\approx0.9$.  We will extend the simulation to
radii $R>2.5 R_*$ to determine whether the restricted size of the
simulation grid has prevented a larger wake from forming.

The photoionization wake does not form at lower latitudes because the disk
prevents X-rays from ionizing the wind in the orbital plane.  The
half-angle of the simulated disk is $\approx8^\circ$, but the X-rays are
shadowed from a half-angle of $20-30^\circ$ by structures in the inner
disk. For all reasonable values of the orbital inclination, the disk shape
determined from the simulation prevents direct X-rays from the neutron
star from being observed at the Earth.  This is consistent with the lack
of X-rays for $\approx0.5$ of the 30~day cycle, but is clearly
inconsistent with the detection of X-ray pulses and a flux implying
near-Eddington luminosities during the rest of the cycle.
In future models, we will improve the spatial resolution of the inner
accretion
disk, to more accurately model the disk structure and X-ray illumination.
This will allow X-rays to be visible at Earth for some disk inclinations,
and may allow a photoionization wake to form over a larger region.

We will also explore the effects of nonsynchronous rotation of the primary
star on the hydrodynamic flow.  The width of optical lines (Hutchings et
al. 1978) suggests the O star rotates at about half of the orbital
frequency.  Levine et al. (1999) measured the orbital period derivative
to be $\dot{P}_{\rm orb}/P_{\rm orb}=(-9.8\pm0.7)\times10^{-7}$~yr$^{-1}$,
and argued, citing similar arguments for the SMC~X-1 system, that this
results from tidal forces operating only for nonsynchronous rotation.
If the primary does not corotate, then the wind may be launched with
an angular velocity (in the rotating frame) that enhances the absorption
at $\phi=0.9$.

We expect that the comparison of global hydrodynamic 
models with observations will continue to test our understanding
of disks, winds, and their interactions.

\appendix
\section{Appendix: The Monte-Carlo Simulation and its Treatment of
Microturbulence}

Our method follows that of Knigge et al. (1995).  We follow randomly
scattering ``photons'' that are actually representative samples of many
photons.  A weight assigned to each ``photon'' determines its relative
contribution to the final profile.  When there is some probability for a
scattering to take place and some probability for the photon to escape
the system,
the photon scatters, but its weight is multiplied by the scattering
probability.  Photons emitted on the stellar surface have some small
probability $\epsilon$ of entering our line of sight.  
As every photon we see must eventually enter our line of sight,
we allow one probability-$\epsilon$ event to occur each time a photon
is emitted or scattered isotropically.
However we no longer follow the path of photons scattered away
from our line of sight, as their contribution should be $\sim \epsilon^2$,
negligibly small.

For each randomly
chosen point on the stellar surface we perform an integration along the
line of sight.  The photon's initial weight is $\mu$, the cosine of the
angle between the surface normal and the line of sight.  When we 
tally the weights of the photons reaching the observer for a given
$\nu$,
we normalize based on the weights $\mu$ that were emitted.
Those photons
that are not initially headed towards the observer are all given a
weight of unity, but the distribution of their random directions 
is chosen to take into account the different fluxes received at
various angles.  At each scattering, there is again a probability of
$\epsilon$ that it enters the line of sight, so we propagage the
photon along the line of sight and collect its weight into 
frequency bins to form the spectrum.  When a photon excites a resonance
line, there is
a probability of $1/[1+n_e q_{21}/A_{21}]$ that the excited atom
will decay through re-emission and not collisions.  Here,
$A_{21}$ is the Einstein A coefficient and
$q_{21}=8.629\times10^{-6} \Sigma / (T^{1/2} g)$ is the decay rate
per ion, with the ``collision strength'' $\Sigma\approx8$ (Clark et al.
1982), and $g$ defined as the statistical weight of the upper level.

Microturbulence affects the Monte-Carlo simulation in two ways.  First,
the probability that a photon is scattered along the line of flight from 
$p$ to $p^\prime$ is $1-\exp(-\tau)$, with 
\begin{equation}
\tau=\int_p^{p^\prime} (\kappa_B +\kappa_R) ds
\end{equation}
Here, $\kappa_B$ ($\kappa_R$) is the scattering coefficient in the blue
(red) doublet component, given by
\begin{equation}
\kappa=\kappa_0 \exp(-\Delta w^2) / \pi \Delta w_T
\end{equation}
where $\Delta w_T=v_T/v_\infty$ is the normalized $1 \sigma$ 
microturbulent velocity and $\Delta w$ is given by
\begin{equation}
\Delta w_B = (c/v_\infty)(\nu-\nu_0)/\nu_0-\mbox{\bf V}\cdot\mbox{\bf
n}
\end{equation}
for the blue doublet component and $\Delta w_R=\Delta w_B + \delta$ 
for the red doublet component.
The factor $\kappa_0$
is set to the optical
depth in Equations~\ref{eqn:empirical} or \ref{eqn:sobolev}, with the
derivatives taken with respect to the line of flight distance $s$ rather
than the radius $r$ from the center of mass of the primary star (radial
optical depths converted to line-of-flight optical depths).
Once it is determined that a photon scatters along a flight line,
the scattering point is determined by the method of Knigge et al.,
and the probability that the photon is scattered by the blue (red)
doublet component is $\kappa_B/\kappa$ ($\kappa_R/\kappa$) where
$\kappa=\kappa_B+\kappa_R$.

Second, a photon's frequency $\nu$ is altered following a scattering,
depending on the microturbulent velocity distribution width $\Delta w_T$.
Given that the macroscopic velocity field has a velocity vector {\bf
V}$(r,\theta,\phi)$ at
point $p$ and that the line of flight is defined by the unit vector {\bf
n}, the actual microturbulent velocity {\bf v$_T$} at point $p$
that scatters the photon must have a component along the line of flight 
{\bf v$_T$}$\cdot${\bf n}$=\Delta w$ (if the red component
has scattered the line then {\bf v$_T$}$\cdot${\bf n}$=\Delta
w+\delta$).
Then the velocities in two directions perpindicular to {\bf n} are given
by $v_\infty \Delta w_T \Phi$, where $\Phi$
is a
random number with Gaussian distribution.  The frequency of
the scattered photon is then $\nu^\prime=\nu_0(1+\mbox{\bf v$_{\rm
wind}$}\cdot\mbox{\bf n}/c)$ for the blue component and
$\nu^\prime=(\nu_0-\delta)(1+\mbox{\bf v$_{\rm wind}$}\cdot\mbox{\bf
n}/c)$ for
the red component.  Here, $\mbox{\bf v$_{\rm
wind}$}= \mbox{\bf V}(r,\theta,\phi)+\mbox{\bf v}_T$ is the actual
velocity
vector of the ion that scatters the photon.

\section{Appendix: Comparison of the Monte Carlo code and the SEI
and Comoving Frame methods}

In this Appendix, we show that our Monte Carlo P~Cygni profile code can
reproduce the results of calculations in the literature using the SEI
method (Lamers et al. 1987) or the Comoving Frame method (Hamann 1981).
We calculate P~Cygni profiles for the wind velocity law
\begin{equation}
w(x)=(1-0.999/x)^{0.5}
\end{equation}
where $w(x)=v(x)/v_\infty$, and
$x=R/R_*$.  We use a uniform microturbulent velocity $\Delta w_T=0.1$ and  
a doublet separation $\delta=(\nu_B-\nu_R)c/\nu_B v_\infty=0.7$.  The
doublet ratio is assumed to be 2:1 (the oscillator strength in the blue
component being greater.)  We make two tests; one for $\tau_B=1$, and
the other for $\tau_B=20$.  The second is a more stringent test, as the
typical photon undergoes more scatterings, and errors introduced at each
scattering are magnified.  For the results of the Comoving Frame method,
we use the values plotted in Fig. 6 of Lamers et al.  For the results of
the SEI method, we use profiles computed with our own version of the SEI
code, which we applied to the LMC~X-4 system in Boroson et al. (1999).

For these tests, we use a spherical star.  In contrast to the case
in which the star fills its Roche lobe, here we know exactly the
fraction of line-of-sight photons that do not intersect the stellar
surface (that is, we know the visible surface area of the star
exactly.)  Thus the continuum level is known, and the number of photons
initially emitted along of the line of sight that arrive at each
frequency is normalized to this continuum level.  The photons that
have scattered into our line of sight are normalized to the total
number emitted by the star.
  
We show the results of the comparison of the three methods in
Figure~\ref{fig:comparison}.  The rms deviation between the SEI and 
Monte Carlo methods is 3\%\ for $\tau=1$ and 7\%\ for $\tau=20$.

\acknowledgements

We would like to thank Daniel Proga, Lex Kaper, Patrick Wojdowski, and
Masao Sako for
useful discussions.
This work was
based on observations with the NASA/ESA {\it Hubble Space Telescope},
obtained at the Space Telescope Science Institute, which is operated by
the Association of Universities for Research in Astronomy, Inc., under
NASA contract GO-05874.01-94A.  
BB acknowledges an NRC
postdoctoral associateship. JMB acknowledges an NSF CAREER Award.

\clearpage

\figcaption{An isodensity surface of the hydrodynamic model at the end
of the numerical simulation, illustrating the relative scales of the
accretion disk, tidal stream and primary star.}

\figcaption{A two-dimensional slice along the line of centers of the
binary system illustrating the five key components of the circumstellar
gas.}

\figcaption{The velocity field of the circumstellar gas in the equatorial
plane
of the binary system.  Here black represents the relatively low density of
the equatorial wind, while white represents the high density of the primary
star, tidal stream, and accretion disk.  The solid line marks the 
critical potential surface.}

\figcaption{The bipolar disk wind emanating from the inner region of the
accretion
disk is evident in this two-dimensional slice along the line of centers of
the binary system.  Note that the disk wind crashes into an irradiation-driven
wind off the surface of the primary star.}

\figcaption{The influence of the disk wind on the circumstellar structure
is
seen in these two-dimensional slices parallel to the equatorial plane.  
The density in the leading wake (left of
the disk wind) is 4 times higher than the normal stellar wind at that 
radius. Note that in the top slice the inner edge of the grid is
beyond the surface of the primary star.}





\figcaption{Simulated \ion{N}{5} P~Cygni profiles from LMC~X-4 (solid
lines),
assuming
density and velocity fields given by hydrodynamic simulations, compared
with P~Cygni profiles observed with the STIS or GHRS (dotted lines).
Velocities are relative to the velocity of the LMC (280 \kms redshift)
and are scaled to a maximum wind velocity of 2000 \kms.
Exposures at $\phi=0.08, 0.28, 0.53,0.68,0.90$ were obtained with the
STIS, while exposures at $\phi=0.11,0.21,0.31,0.41$ were obtained with the
GHRS}

\figcaption{Simulated \ion{C}{4} P~Cygni profiles from LMC~X-4 (solid
lines), assuming
density and velocity fields given by hydrodynamic simulations, compared
with spectra observed with the STIS or GHRS (dotted lines).
Exposures
at $\phi=0.08, 0.28, 0.53, 0.68, 0.90$ were obtained with the STIS, while
exposures at $\phi=0.16,0.26,0.36,0.46$ were obtained with the GHRS.}

\figcaption{Examination of the orbital asymmetry in the \ion{N}{5} line
profiles.
From X-ray ionization alone, the profiles at $\phi,1-\phi$ should be
identical.  The top left panel shows that the profile at $\phi=0.08$
(solid line) does not have as much absorption as the profile at
$\phi=0.90$ (dotted line).  The bottom left panel compares the
profiles at $\phi=0.28$ (solid line) and $\phi=0.68$ (dotted line).
Panels on the right show the predictions from the simulations.}

\figcaption{Differences between \ion{N}{5} profiles predicted from
the simulation at two time frames.  The dotted line shows the
expected error from the simulated counting statistics.}

\figcaption{The X-ray spectrum expected from the wind in mid-eclipse
in the simulation and as observed with ASCA.  The error bars and limits 
show the ASCA data, while the Compton-scattered, bremsstrahlung,
recombination line, and fluorescent line spectra folded with the ASCA
response are indicated by ``Compt'', ``Brems'', ``Recom'', and
``Fluor''.}


\figcaption{A comparison of the results of our Monte-Carlo P Cygni line
simulation (solid line) with the results of the SEI method ($+$ signs) and
Comoving Frame method ($\Diamond$ signs.)  Top panel: optical depths
$\tau_B=20, \tau_R=10$.  Bottom panel: $\tau_B=1$, $\tau_R=0.5$.}

\clearpage

\begin{table*}
\begin{center}
\begin{tabular}{llcc}
Parameter name & Meaning & Value & Reference\\
\hline
$L_{\rm x}$ & X-ray luminosity & $2\times10^{38}$~erg~s$^{-1}$ & *\\
$M_{\rm O}$ & Mass of O star & 14.6$\Msun$ & *(4) \\
$M_{\rm ns}$ & Mass of neutron star & 1.4$\Msun$ & *(4)\\
$P_{\rm orbit}$ & Orbital period & 1.4~d & 3\\
$i$ & Orbital inclination & $65^\circ$ & *(3,4)\\
$a$ & Separation of centers of mass & $9.35\times10^{11}$ cm & *(3,4)\\
  & Spectral type & O\,7III-V & 2\\
$a_{\rm N}$ & Nitrogen abundance & $10^{-4}$ & *\\
$a_{\rm C}$ & Carbon abundance & $3.6\times10^{-4}$ & *\\
$\gamma$ & Wind depth exponent & 0.75 & 1 \\
$T$ & Total wind depth & 1.15 & 1\\
$v_{\rm T}/v_\infty$ & Turbulent velocity & 0.04 & *\\
\end{tabular}
\end{center}
\caption{Adopted parameters of the LMC~X-4 system}
$^*$Adopted\\
$^1$Boroson et al. 1999\\
$^2$Hutchings, Crampton, \&\ Cowley 1978\\
$^3$Kelley et al. 1981\\
$^4$Levine et al. 1991\\
\end{table*}

\clearpage

\setcounter{figure}{0}

\begin{figure}
\caption{See JPEG file}
\end{figure}

\begin{figure}
\caption{See JPEG file}
\end{figure}

\begin{figure}
\caption{See JPEG file}
\end{figure}

\begin{figure}
\caption{See JPEG file}
\end{figure}

\begin{figure}
\caption{See JPEG file}
\end{figure}

\newpage

\begin{figure}
\caption{\label{fig:n5frame1}}
\plotone{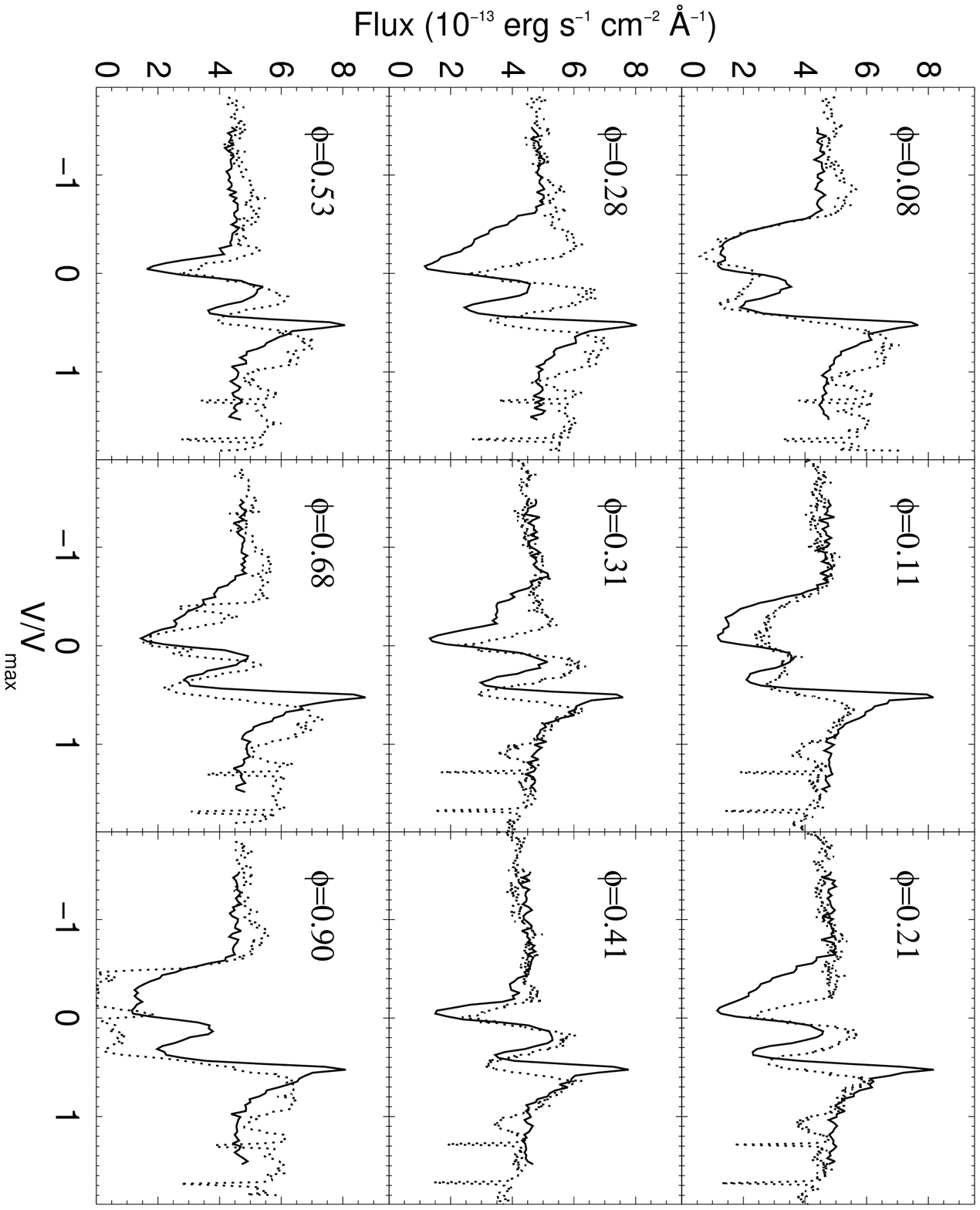}
\end{figure}

\begin{figure}
\caption{\label{fig:c4frame1}}
\plotone{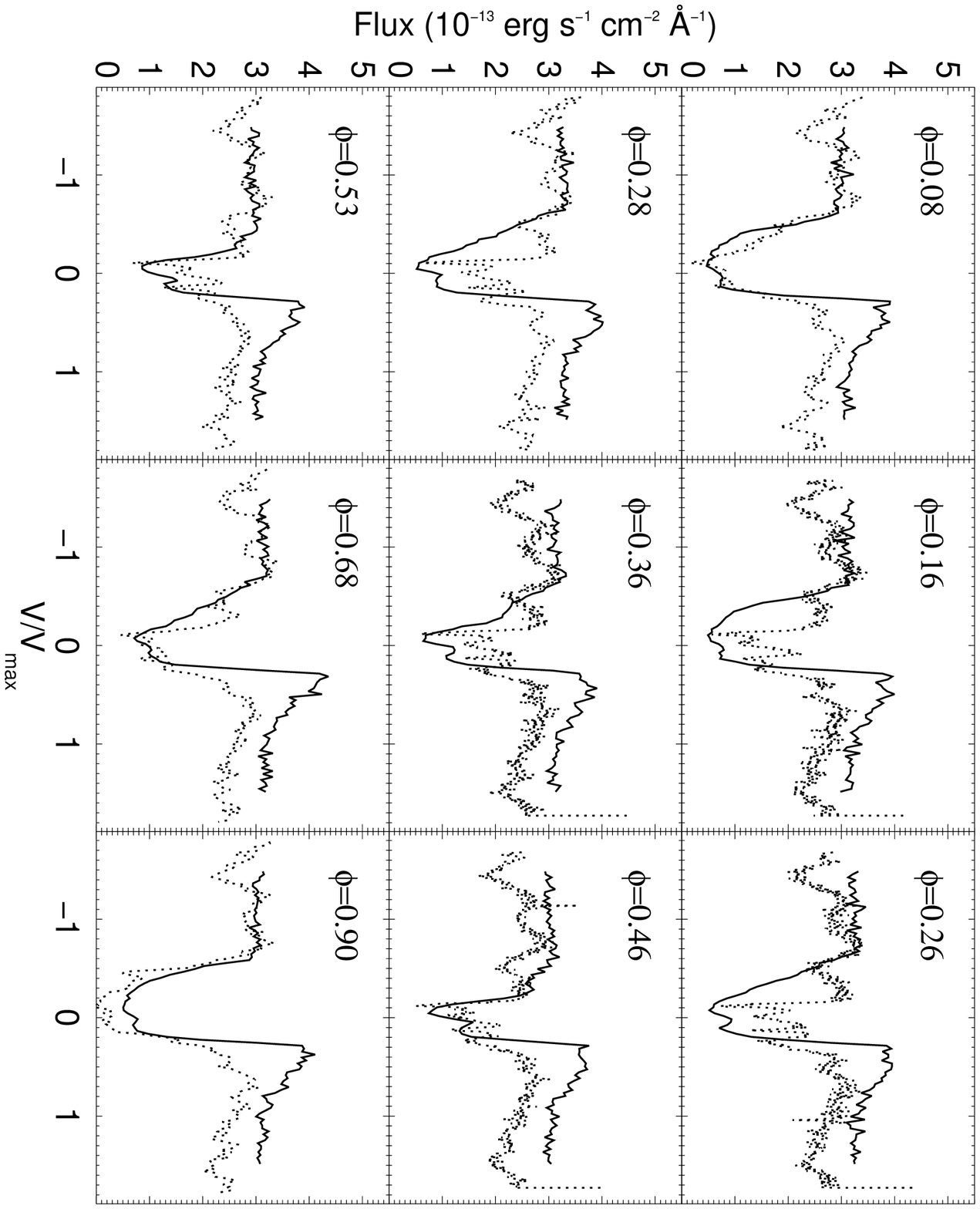}
\end{figure}

\begin{figure}
\caption{\label{fig:asym}}
\plotone{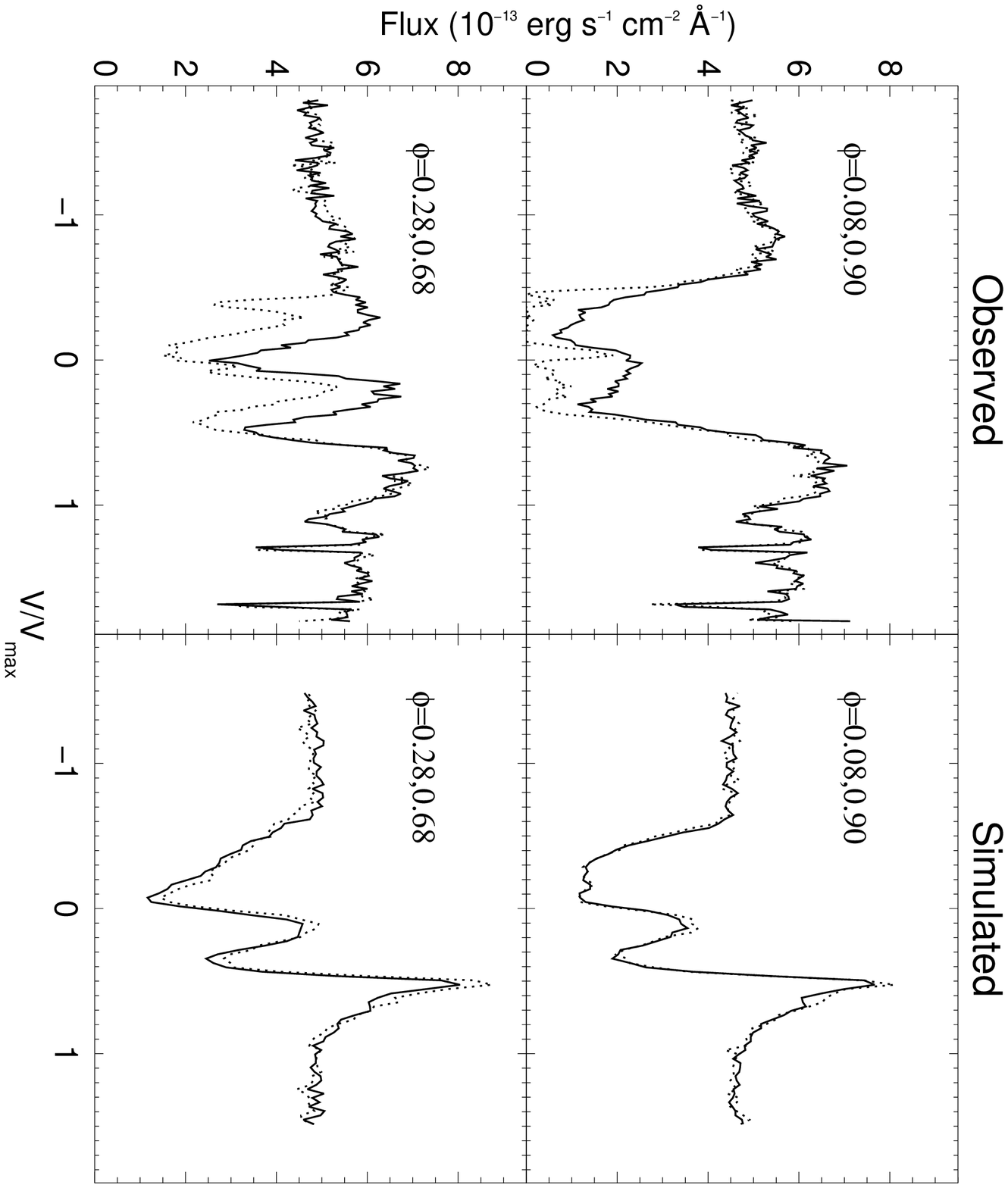}
\end{figure}

\begin{figure}
\caption{\label{fig:difs}}
\plotone{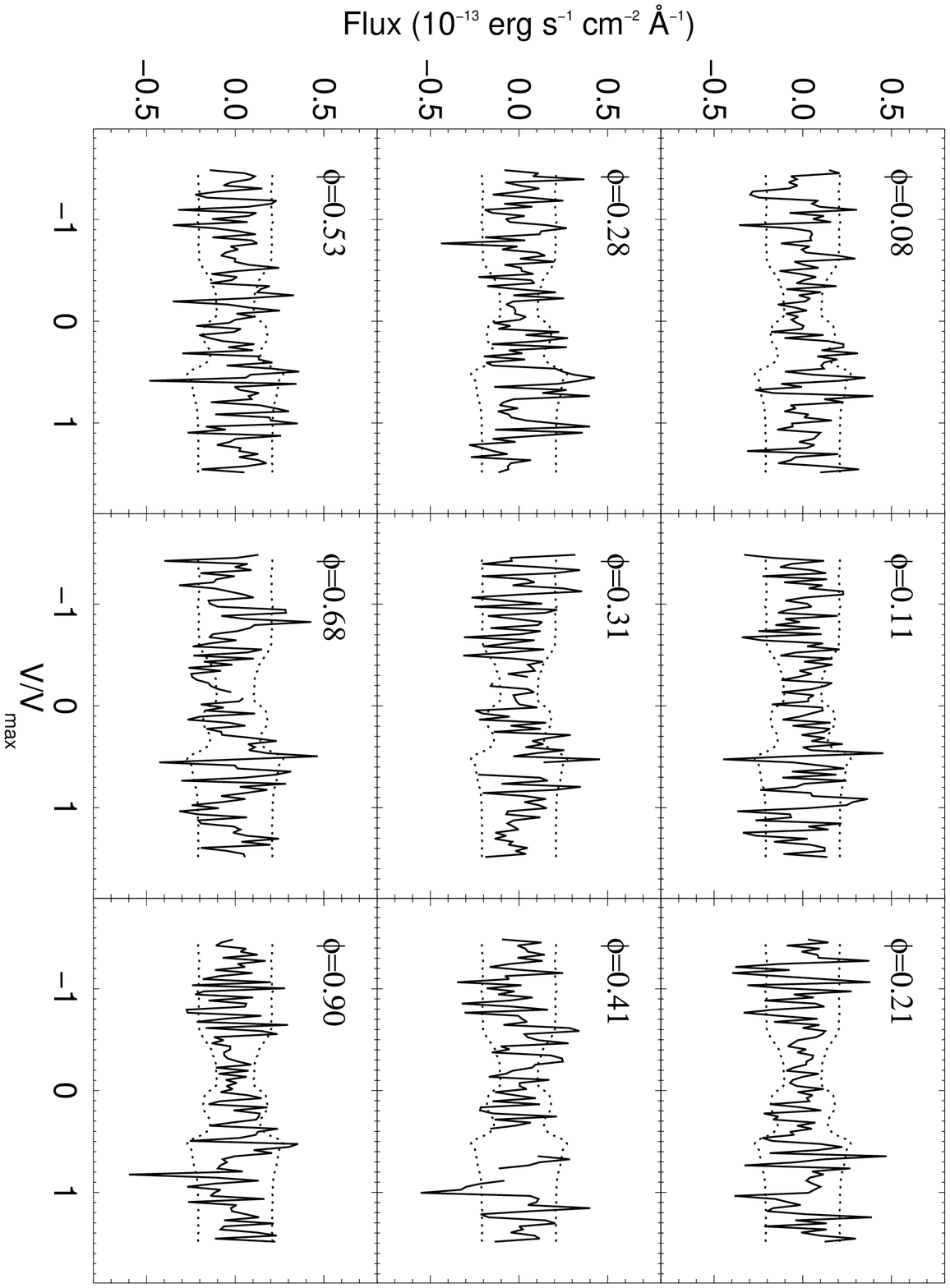}
\end{figure}

\begin{figure}
\caption{\label{fig:windemit}}
\plotone{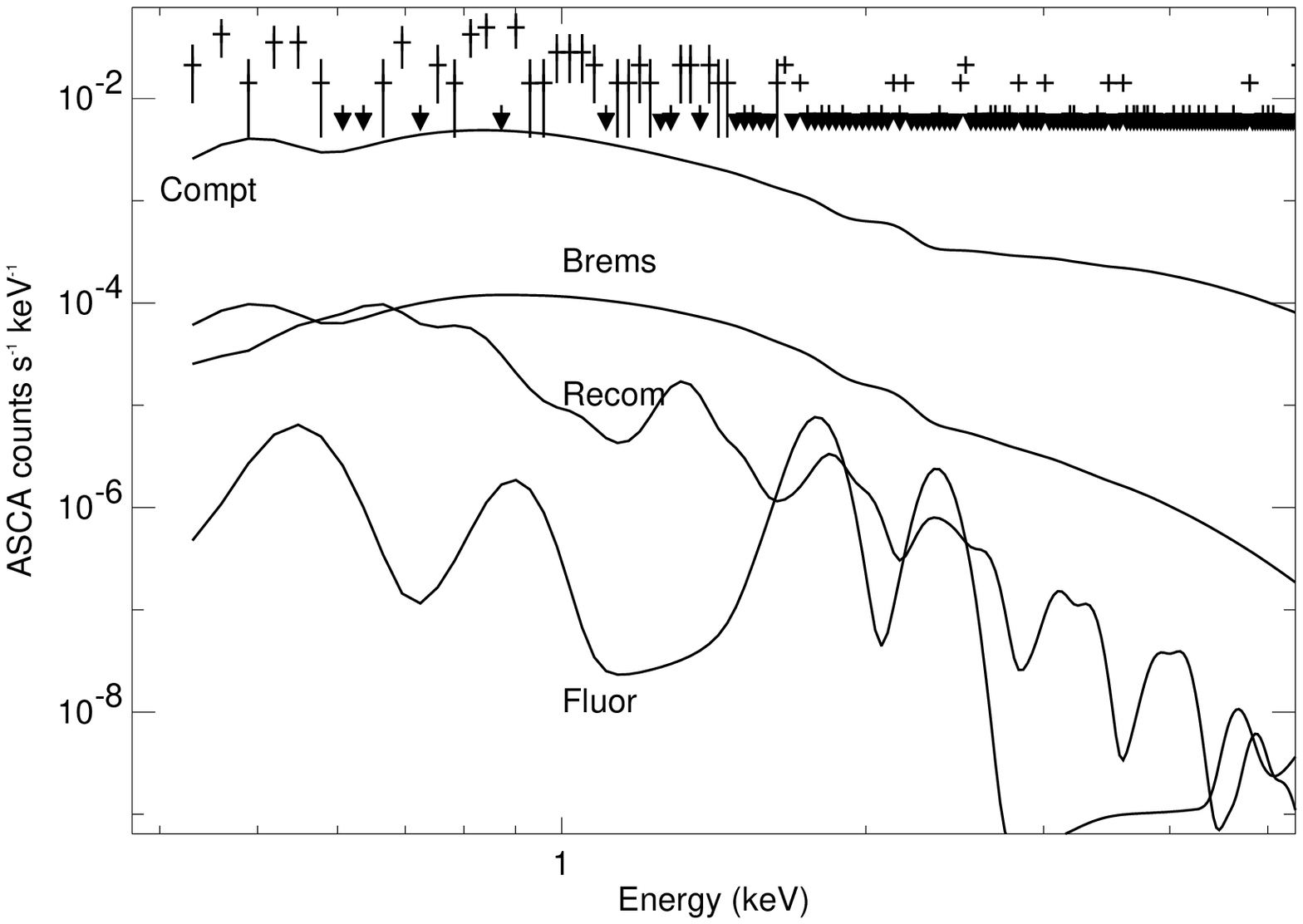}
\end{figure}

\begin{figure}
\caption{\label{fig:comparison}}
\plotone{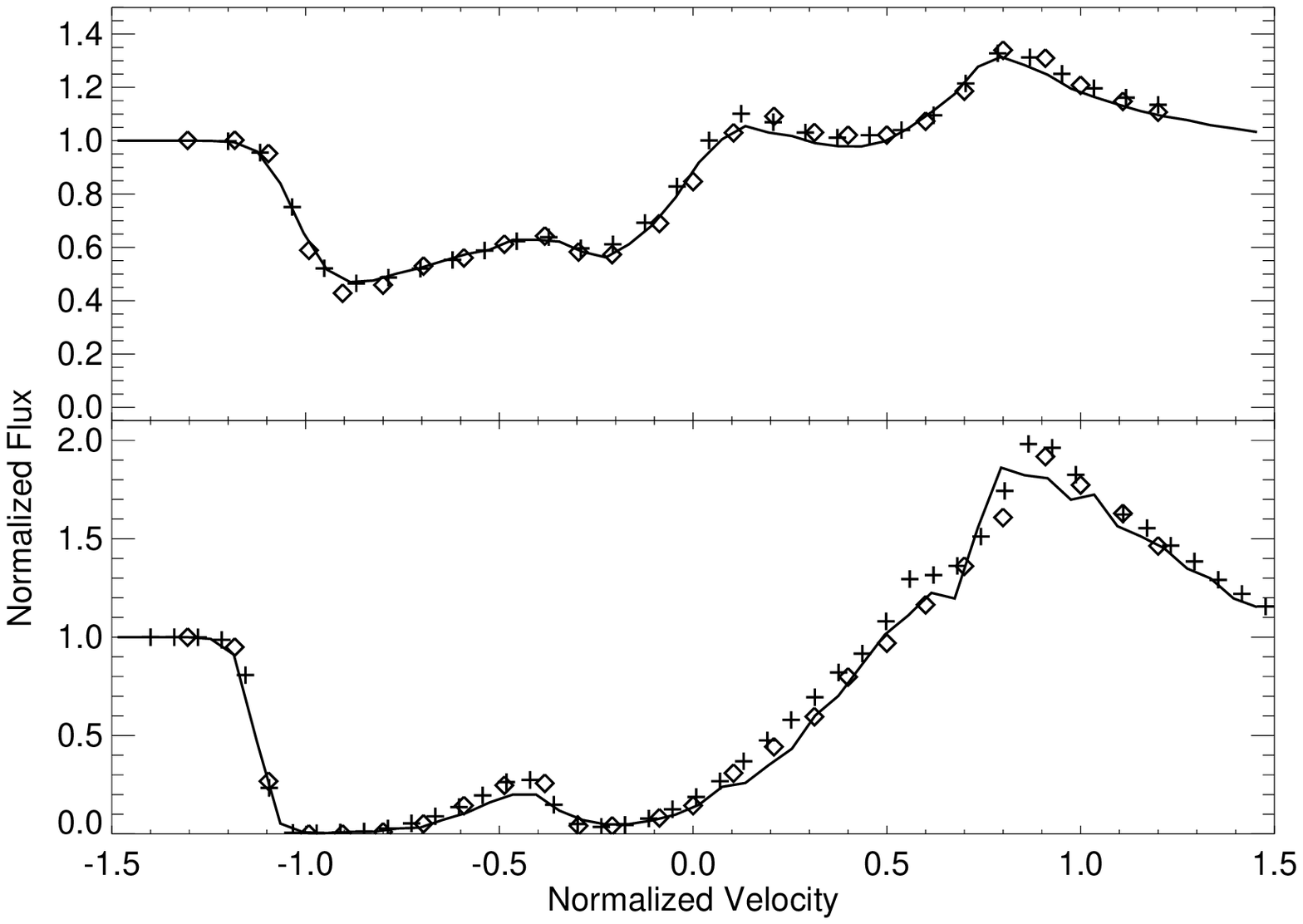}
\end{figure}

\end{document}